\documentclass{iopart}
 \newcommand{\be}[1]{\begin{equation}\label{#1}}
 \newcommand{\ba}[1]{\begin{eqnarray}\label{#1}}

 \newcommand{\de}[1]{\delta_{#1}}
 \newcommand{\pd}[1]{\frac{\partial}{\partial #1}}
 \newcommand{\ppd}[1]{\frac{\partial^2}{\partial {#1}^2}}

 \newcommand{\re}{{\rm e}}
 \newcommand{\pa}[1]{\left(#1\right)}
 \newcommand{\paq}[1]{\left[#1\right]}
 \newcommand{\pag}[1]{\left\{#1\right\}}
 \newcommand{\av}[1]{\langle#1\rangle}

 \def\ee{\end{equation}}
 \def\ea{\end{eqnarray}}

\usepackage{graphicx,latexsym}
\usepackage{graphicx,epsf}
\usepackage{color}
\usepackage{epsfig}
\usepackage[T1]{fontenc}
\usepackage[utf8]{inputenc}
\usepackage{tikz}
\usepackage{pgfplots}

\begin{document}
\title[The Born-Oppenheimer Method, Quantum Gravity and Matter]{\textbf{The Born-Oppenheimer Method, Quantum Gravity and Matter}}

\author{Alexander~Yu~Kamenshchik$^{1,2}$, Alessandro~Tronconi$^{1}$,
Giovanni~Venturi$^{1}$}
\address{$^1$Dipartimento di Fisica e Astronomia and INFN, \small Via Irnerio 46, 40126, Bologna,
Italy,\\
$^2$L.D. Landau Institute for Theoretical Physics of the Russian
Academy of Sciences, Kosygin str. 2, 119334, Moscow, Russia}
\eads{\mailto{Alexander.Kamenshchik@bo.infn.it},
 \mailto{Alessandro.Tronconi@bo.infn.it}, \mailto{Giovanni.Venturi@bo.infn.it}}
\begin{abstract}
We illustrate and examine diverse approaches to the quantum matter-gravity system which refer to the Born-Oppenheimer (BO) method. In particular we first examine a quantum geometrodynamical approach introduced by other authors in a manner analogous to that previously employed by us, so as to include back reaction and non-adiabatic contributions. On including such effects it is seen that the unitarity violating effects previously found disappear. A quantum loop space formulation (based on a hybrid quantisation, polymer for gravitation and canonical for matter) also refers to the BO method. It does not involve the classical limit for gravitation and has a highly peaked initial scalar field state. We point out that it does not resemble in any way to our traditional BO approach. Instead it does resemble an alternative, canonically quantised, non BO approach which we have also previously discussed. 
\end{abstract}
\section{Introduction}
Quantum cosmology, i.e. the treatment of the universe as a unique quantum object, governed by the laws of general relativity and quantum field theory, is becoming a recognised part of modern theoretical physics. This is connected with at least three reasons. Firstly, treatment of the universe as a quantum object is a natural part of the program of the unification of all the fundamental interactions, including gravitational ones. Secondly, the success of inflationary cosmology \cite{inflation} requires a further investigation of the physics of the very early universe. Thirdly, the mathematical structure of quantum gravity and cosmology is close to that of such popular theories as string and superstring models. Indeed, it represents a theory with first class constraints, having reparametrisation invariance, or invariance with respect to the spacetime diffeomorphisms. The main goal of quantum cosmology is the description of the quantum state of the universe. Such a state should satisfy the Wheeler-DeWitt equation (WDW) \cite{DeWitt}, which arises as a result of the application of the Dirac quantisation procedure \cite{Dirac} to the universe. The study of the quantum state of the universe and the Wheeler-DeWitt equation encounters serious difficulties. Firstly, it involves a huge number of degrees of freedom and is enormously cumbersome. Thus, the application of some approximation schemes is necessary. On then working with the constrained theory one should be able to choose and separate the physical degrees of freedom from the gauge ones. Even if the wave function of the universe is then constructed, its probabilistic interpretation is not immediate: one should give the appropriate structure to the corresponding Hilbert state \cite{Barv,Barv-Kam}. Since the natural notion of time is absent in quantum cosmology, one has to identify with time some combination of degrees of freedom. Moreover, different choices of boundary conditions for the Wheeler-DeWitt equation are possible. The most famous tunneling \cite{Vilenkin} and no-boundary \cite{Hartle} cosmological wave functions give different physical predictions. 

Historically, the first models for which some solutions of the Wheeler-DeWitt equation were constructed were the so called minisuperspace models, where only a small finite 
number of degrees of freedom was taken into account. These degrees of freedom were connected with the global characteristics of the universe. The next natural step in the development of quantum cosmology is the quantisation of cosmological perturbations. At this stage, one can start with some global degrees of freedom and on this background treat the inhomogeneous harmonics at tree level as in \cite{Hall-Hawk} or also consider one-loop quantum corrections to the tree-level wave function of the universe \cite{Barv-Kam1,Barv-Kam2}. The latter approach allows one to obtain the normalised wave function of the universe \cite{Barv-Kam1} and predict the most probable initial conditions for inflation \cite{Barv-Kam2}. 

The separation between the background variables and the cosmological perturbations both for the gravitational degrees of freedom and for the matter ones, used in the papers mentioned above, was based on the purely geometrical characteristics. Namely, the spatially homogeneous components of the metric and of the matter fields were taken as background variables, while the inhomogeneous ones were treated as perturbations. Such an approach descends from classical cosmology \cite{Lif-Khal}. 
However, another approach to the treatment of cosmological degrees of freedom is possible. The point is that in reality we have two different mass/time scales
in cosmology. One is connected with the Planck mass (Planck length, Planck time) and characterises gravitation, the other scale is connected with the non-gravitational matter fields. This fact can direct one's attention to a well-known method for the treatment of molecules such as the Born-Oppenheimer (BO) approach \cite{BO}. 

Composite systems which involve two mass (or time) scales such as molecules are amenable to this particular treatment. For molecules this is possible because of the different nuclear and electron masses, which allows one to suitably factorise the wave function of the composite system leading, in a first approximation, to a separate description of the motion of nuclei and electrons. In particular it is found that the former are influenced by the mean Hamiltonian of the latter and the latter (electrons) follow the former adiabatically (in the quantum mechanical sense). By this we mean that the nuclei move sufficiently slowly so that the electrons do not change eigenstates. Clearly should the nuclei move quickly enough, the electrons will undergo (non-adiabatic) transitions between states.

It has been pointed out that the matter-gravity system is also amenable to such an approach \cite{Brout} since gravity is characterised by the Planck mass, which is much greater than the usual matter mass, then the heavy degrees of freedom are associated with gravitation and the light ones with matter. As a consequence, to lowest order, gravitation will be driven by the average matter Hamiltonian and matter should follow gravity adiabatically. The original motivation for such an approach \cite{Brout} was to examine the emergence of time in the context of quantum gravity where time does not appear as a consequence of reparametrisation invariance. This then led to a study of the semiclassical limit for gravitation starting from a mini-superspace formulation for quantum gravity \cite{DeWitt} and quantum matter. One then obtained the Einstein equations with gravity driven by the (quantum) mean energy Hamiltonian of matter \cite{Brout}.

Subsequently the approach was further analysed in order to justify the use of the mean matter energy to drive gravitation \cite {GV}. This was found to be true when non-adiabatic transitions (fluctuations) are negligible and the Universe is sufficiently far from the Planck scale. This occurs during inflation and one then has the usual (unitary) time evolution of quantum matter (Schwinger-Tomonaga or Schr\"odinger).

Initially the BO approach in cosmology addressed mini-superspace and homogeneous matter modes. Subsequently it was generalised to metric perturbations and non-homogeneous modes in order to obtain corrections to the usual power spectrum of cosmological fluctuations produced during inflation \cite{T}. These corrections, which essentially amount to the inclusion of the non-adiabatic transitions, can affect the infrared part of the spectrum and may lead to an amplification or suppression depending on the background evolution.
Indeed, in all the above our ultimate scope was to obtain corrections to the usual power spectrum of cosmological fluctuations produced during inflation.

Other authors have applied the BO approach (or what they call a BO approach), again with the aim of finding its effect on the power spectrum. In particular, in the next sections we illustrate two diverse approaches both involving gravitation and matter. 

The first geometrodynamical approach by Brizuela, Kiefer and Kramer (henceforth BKK) \cite{KieferdS,Kiefer} is examined in detail in Section 2, in particular with the aim of comparing it with our, above mentioned, traditional BO method which, in contrast, does not suffer from unitarity violating difficulties.

The second approach \cite{lqc} employs a (polymer quantised) loop space formulation for gravitation together with a Fock space quantisation for matter. Concerning this approach in Section 3 we shall limit ourselves to a brief description of it and observe that even if it has nothing in common, both technically and physically, with our first, traditional BO method, it does bear some resemblance with an alternative 
(non BO) approach which we have previously examined \cite{TVV}. It consisted of the study of a quantum matter-gravity system containing a minimally coupled massive homogeneous scalar field which is known to lead to inflation. After choosing a suitable initial state for the scalar field, the equations for the homogeneous gravity matter system is solved in the inflationary (scale factor $a$ large) limit. On then introducing other matter fields (or inhomogeneous modes), after coarse graining of the gravitational wave function, an effective time evolution emerges for them. In this case the presence of an effective time evolution for matter arises from a mechanism similar to one already observed in the analysis of the classical limit of quantum systems, such as the hydrogen atom \cite{rowe}. In particular the zero angular momentum and large principal quantum number case which exhibits a radial highly oscillatory behaviour. In this case on coarse graining (in particular on applying the Riemann-Lebesgue Lemma) one is able to recover the classical trajectory. 
Indeed, the classical trajectory is related to a classical spatial probability distribution of a particle in terms of the inverse of its speed (the fraction of time spent in a spatial interval is a measure of the probability density). There is a deep connection between the above example and the situation present in the matter-gravity system for this case.

Lastly our results are summarised and discussed in the conclusions. 

\section{Geometrodynamical Approach}
The BKK approach \cite{KieferdS} is based on the Wheeler-de Witt equation and can be directly compared with ours \cite{T}. Indeed, in \cite{KieferdS}, they introduce a ``master'' Wheeler-de Witt equation (we shall follow an identical notation to theirs whenever possible) obtained after a Fourier expansion (we work on a flat FRW background): 
\be{wdw1}
\!\!\!\!\!\!\!\!\!\!
\!\!\!\!\!\!\!\!\!\!
\!\!\!\!\!\!\!\!\!\!
\!\!\!\!\!
\frac{1}{2}\paq{-\frac{\hbar^2}{m^2}\mathcal{G}^{AB}\frac{\partial^2}{\partial q^A\partial q^B}+m^2 V(q)+\sum_k\pa{-\hbar^2\ppd{v_k}+\omega_k^2v_k^2}
}\Psi\pa{q,\{v_k\}}=0,
\ee
where the minisuperspace variables $q^2\equiv \alpha=\ln\pa{a/a_0}$, the logarithm of the scale factor,  $q^1\equiv \phi/m$, which is a homogeneous scalar field $\phi$ scaled by the Planck mass $m$, have been introduced. In the above one has a metric $\mathcal{G}^{AB}\equiv {\rm diag}\pa{-\re^{-2\alpha},\re^{-2\alpha}}$ and a potential $V(q)$ with $q=\pag{q^1,q^2}$. Further $\omega_k(q)\equiv \omega_k(\eta)$ where $\eta$ is the conformal time and $v_k$ is the Mukhanov-Sasaki (MS)  \cite{Mukhanov,Sasaki} variable associated with the wave number $k$. 
On following a BO approach \cite{BO} we expand $\Psi$ onto a complete basis of states for the matter field \footnote{A suitable lowest order basis could be the solutions to the time dependent harmonic oscillator equations of motion \cite{Lewis,Bertoni}.} and subsequently separate the equations of motion for the homogeneous (minisuperspace) and inhomogeneous parts of the wave function:
\be{WFdec}
\Psi(q,\pag{v_k})=\Psi_0\pa{q}\prod_k\tilde{\Psi}_k\pa{q,v_k}\equiv \Psi_0(q)\tilde \Psi\pa{q,\pag{v_k}}
\ee
where the tilde just labels the parts of the wave function containing the inhomogeneous variables.
We emphasise that hereafter, in contrast with our approach \cite{T}, the homogeneous mode for the scalar field is included in the ``slow'' part of the factorisation. This occurs since the scalar field has been rescaled with $m$ and has consequences on the Einstein equations obtained. This is somewhat artificial and in contrast with the spirit of the BO approach which we shall nonetheless follow.\\
On multiplying the above by $\tilde \Psi^*$ and integrating over all $v_k$ one obtains the equation for the gravitational part of the total wave function
\be{wdwG}
\!\!\!\!\!\!\!\!\!\!
\!\!\!\!\!\!\!\!\!\!
\!\!\!\!\!\!\!\!\!\!
\!\!\!\!\!\!\!\!
\paq{-\frac{\hbar^2}{m^2}\mathcal{G}^{AB}\frac{\partial^2}{\partial q^A\partial q^B}+m^2 V(q)+\sum_k\av{\hat H_k}}\bar\Psi_0\pa{q}=\frac{\hbar^2}{m^2}\mathcal{G}^{AB}\av{\frac{\partial^2}{\partial q^A\partial q^B}}\bar\Psi_0\pa{q}, 
\ee
where 
\be{geophi0}
\Psi_0(q)=\bar{\Psi}_0(q)\exp\paq{i\int^{q}\mathcal{A}_A(Q) d{Q}^A},
\ee
\be{defA}
\mathcal{A}_A(q)\equiv i \av{\tilde\Psi|\frac{\partial}{\partial q^A}|\tilde \Psi}=i\int  {\tilde\Psi^*(q,\pag{v_k})} \frac{\partial }{\partial q^A}\tilde\Psi(q,\pag{v_k})\prod_k dv_k
\ee
and $Q^A$ is a dummy integration variable,
\be{Hkdef}
\hat H_k=\frac{1}{2}\paq{-\hbar^2\ppd{v_k}+\omega_k(\eta)^2v_k^2},
\ee
 with 
\be{avdef}
\av{\hat O}\equiv \int  {\bar{\tilde\Psi}}^*\pa{q,\pag{v_k}}\hat O\, {\bar{\tilde\Psi}}\pa{q,\pag{v_k}}\prod_k dv_k
\ee
and
\be{geophitilde}
\tilde\Psi\pa{q,\pag{v_k}}=\bar{\tilde\Psi}\pa{q,\pag{v_k}}\exp\paq{-i\int^{q}\mathcal{A}_A(Q) dQ^A},
\ee
where each matter mode is individually normalised and $Q^A$ is a dummy integration variable. Actually all the above can be repeated using non normalised states without any change in the results, this is done in \cite{Bertoni,Venturi90}, here we have just used normalised states for simplicity. Let us note that the barred wave functions introduced by (\ref{geophi0}) and (\ref{geophitilde}) are defined just through a $q$ dependent phase factorisation which is opposite for (\ref{geophi0}) and (\ref{geophitilde}) in order to leave their product phase invariant.

As before \cite{T} one can obtain an equation for the matter wave function $\bar{\tilde\Psi}(q,\pag{v_k})$ and, in particular, project out a single Fourier component $\bar{\tilde\Psi}_k\pa{q,v_k}$ obtaining
\ba{mateq0array}
&&\paq{\hat H_k-\av{\hat H_k}_k}\bar{\tilde\Psi}_k-\frac{\hbar^2}{m^2}\mathcal{G}^{AB}\pd{q^A}\ln \bar\Psi_0\pd{q^B}\bar{\tilde\Psi}_k\nonumber\\
&&=\frac{\hbar^2}{2m^2}\mathcal{G}^{AB}\paq{\frac{\partial^2}{\partial q^A\partial q^B}-\av{\frac{\partial^2}{\partial q^A\partial q^B}}_k}\bar{\tilde\Psi}_k,\label{mateq0}
\ea
where 
\be{avkdef}
\av{\hat O}_k\equiv \int dv_k {\bar{\tilde\Psi}}_k^*(q,v_k)\hat O \,{\bar{\tilde\Psi}}_k(q,v_k).
\ee
The equations (\ref{wdwG}) and (\ref{mateq0}) are equivalent to (\ref{wdw1}) and are the expected outcome of the BO decomposition.\\ 
BKK, in \cite{KieferdS}, in order to obtain the Schr\"odinger-like equation for the wave function of the inhomogeneous matter modes, the Eq. (\ref{wdw1}) is surprisingly (since the new equation is incompatible with the previous one, however see later) replaced by an equation with a single mode:
\be{wdwk}
\!\!\!\!\!\!\!\!\!\!
\!\!\!\!\!\!\!\!\!\!
\frac{1}{2}\paq{-\frac{\hbar^2}{m^2}\mathcal{G}^{AB}\frac{\partial^2}{\partial q^A\partial q^B}+m^2 V(q)-\hbar^2\ppd{v_k}+\omega_k^2v_k^2
}\Psi_k\pa{q,v_k}=0.
\ee
If we apply the BO decomposition to (\ref{wdwk}) and define
\be{defPsik}
\Psi_k\pa{q,v_k}\equiv \Psi_0\pa{q}\tilde{\Psi}_k\pa{q,v_k}
\ee
instead of Eq. (\ref{wdwG}) one now obtains from Eq. (\ref{wdwk}) the following equation for the homogeneous part
\ba{wdwGkarray}
&&\paq{-\frac{\hbar^2}{m^2}\mathcal{G}^{AB}\frac{\partial^2}{\partial q^A\partial q^B}+m^2 V(q)+\av{\hat H_k}_k}\bar\Psi_0\pa{q}\nonumber\\
&&=\frac{\hbar^2}{m^2}\mathcal{G}^{AB}\av{\frac{\partial^2}{\partial q^A\partial q^B}}_k\bar\Psi_0\pa{q}\label{wdwGk}, 
\ea
where now 
\be{geophi0k}
\Psi_0(q)=\bar{\Psi}_0(q)\exp\paq{i\int^{q}\mathcal{A}_{A,k}\!\pa{Q}\,dQ^A}
\ee
with
\be{defAk} 
\mathcal{A}_{A,k}(q)\equiv i \av{\tilde\Psi_k|\frac{\partial}{\partial {q}^A}|\tilde\Psi_k}=i\int dv_k {\tilde\Psi_k^*(q,v_k)} \frac{\partial}{\partial q^A}\tilde\Psi_k(q,v_k)
\ee 
and $Q^A$ is a dummy integration variable.
Le us note that, if one just considers one mode, the $\bar{\Psi}_0$ in Eqs. (\ref{wdwG}) and (\ref{wdwGk}) differ unless the Hamiltonian of the perturbations is negligible.\\ 
Concerning the inhomogeneous matter modes, in this case one does not have to perform a projection onto a particular mode and one immediately obtains (\ref{mateq0}). The equations (\ref{wdwGk}) and (\ref{mateq0}) obtained from the BO decomposition are equivalent to (\ref{wdwk}).

Let us follow the approach of BKK in Ref. \cite{KieferdS} and, instead of the BO factorisation (\ref{defPsik}), we make the ansatz 
\be{wkb0}
\Psi_k(q,v_k)=\re^{\frac{i}{\hbar}S(q,v_k)}
\ee
with
\be{Sexp}
S=m^2S_0+m^0 S_1+m^{-2}S_2+\dots
\ee
On substituting the previous expression in Eq. (\ref{wdwk}) and collecting different powers of $m$ one has
\ba{Omser}
&\!\!\!\!\!\!\!\!\!\!\!\!\!\!\!\!\!\!\!\!\!\!\!\!\mathcal{O}\pa{m^4}: &\quad\pd{v_k}S_0=0\label{Om4},\\
&\!\!\!\!\!\!\!\!\!\!\!\!\!\!\!\!\!\!\!\!\!\!\!\!\mathcal{O}\pa{m^2}: &\quad \mathcal{G}^{AB}\frac{\partial S_0}{\partial q^A}\frac{\partial S_0}{\partial q^B}+V=0\label{Om2}.
\ea
We observe that if one identifies $\Psi=\re^{\frac{i}{\hbar}S}$, with $S$ given by (\ref{Sexp}), and substitutes in Eq. (\ref{wdw1}) the results (\ref{Om4}) and (\ref{Om2}) are unchanged since $H_k$ is $\mathcal{O}(m^0)$.\\ Continuing with Eq. (\ref{wdwk}), one has 
\ba{Omserc}
&\!\!\!\!\!\!\!\!\!\!\!\!\!\!\!\!\!\!\!\!\!\!\!\!\mathcal{O}\pa{m^0}: &\quad 2\mathcal{G}^{AB}\frac{\partial S_0}{\partial q^A}\frac{\partial S_1}{\partial q^B}-i\hbar \mathcal{G}^{AB}\frac{\partial^2S_0}{\partial q^A\partial q^B}-i\hbar \frac{\partial^2 S_1}{\partial v_k^2}+\omega_k^2v_k^2=0,
\label{Om0}\\
&\!\!\!\!\!\!\!\!\!\!\!\!\!\!\!\!\!\!\!\!\!\!\!\!\mathcal{O}\pa{m^{-2}}: &\quad \mathcal{G}^{AB}\frac{\partial S_0}{\partial q^A}\frac{\partial S_2}{\partial q^B}+\frac{1}{2}\mathcal{G}^{AB}\frac{\partial S_1}{\partial q^A}\frac{\partial S_1}{\partial q^B}, \nonumber\\
&&\quad-\frac{i\hbar}{2} \mathcal{G}^{AB}\frac{\partial^2 S_1}{\partial q^A\partial q^B}+ \frac{\partial S_1}{\partial v_k}\frac{\partial S_2}{\partial v_k}-\frac{i\hbar}{2} \frac{\partial^2 S_2}{\partial v_k^2}=0.\label{Om-2}
\ea
It is clear that had we used Eq. (\ref{wdw1}) we would have had a $\sum_k$ in Eqs. (\ref{Om0}) and (\ref{Om-2}). The above approach followed by BKK in \cite{KieferdS} is what the authors also call a ``BO scheme'' because, on collecting the different powers of $m$, the total wave-function is splitted into a minisuperspace (homogeneous) part satisfying Eq. (\ref{Om2}) and an inhomogeneous part of higher order in $m$.

Ley us perform an $m$ expansion analogous to (\ref{Sexp}) of Ref. \cite{KieferdS} in our Eqs. (\ref{wdwGk}) and (\ref{mateq0}) which have been obtained through a BO factorisation. We also use the identification 
\ba{wkbsplitarray}
\Psi_k&=&\Psi_0\tilde{\Psi}_k=\bar{\Psi}_0\bar{\tilde{\Psi}}_k=\gamma^{-1}\re^{m^2\frac{i}{\hbar}S_0+m^{-2}\frac{i}{\hbar}\zeta}\;\gamma\,\re^{\frac{i}{\hbar}S_1+m^{-2}\frac{i}{\hbar}\chi}\nonumber\\
&\equiv&\gamma^{-1}\re^{m^2\frac{i}{\hbar}S_0+m^{-2}\frac{i}{\hbar}\zeta}\psi_k^{(1)},\label{wkbsplit}
\ea
where we have further decomposed $S_2$ of (\ref{Sexp}) into $S_2=\zeta(q)+\chi(q,v_k)$ and $\gamma$ is a prefactor (related to the Van Vleck determinant) which is associated with the WKB approximation and is a function of q. If one now substitutes in Eqs. (\ref{wdwGk}), on keeping terms to different orders in $m$ one obtains:
\ba{wkbsplitexp}
&\!\!\!\!\!\!\!\!\!\!\!\!\!\!\!\!\!\!\!\!\!\!\!\!\!\!\!\!\!\!\!\!\!\!\!\!\!\!\mathcal{O}\pa{m^2}:&\!\!\!\!\!\!\!\!\!\!\!\!\quad \mathcal{G}^{AB}\frac{\partial S_0}{\partial q^A}\frac{\partial S_0}{\partial q^B}+V=0\,\label{spOm2}\\
&\!\!\!\!\!\!\!\!\!\!\!\!\!\!\!\!\!\!\!\!\!\!\!\!\!\!\!\!\!\!\!\!\!\!\!\!\!\!\mathcal{O}\pa{m^0}:&\!\!\!\!\!\!\!\!\!\!\!\!\quad \frac{\hbar}{2}\mathcal{G}^{AB}\paq{\frac{2i}{\gamma^2}\frac{\partial \gamma}{\partial q^A}\frac{\partial S_0}{\partial q^B}-\frac{i}{\gamma}\frac{\partial^2 S_0}{\partial q^A \partial q^B}}+\frac{\av {\hat H_k}_0}{\gamma}=0,\label{spOm0}\\
&\!\!\!\!\!\!\!\!\!\!\!\!\!\!\!\!\!\!\!\!\!\!\!\!\!\!\!\!\!\!\!\!\!\!\!\!\!\!\mathcal{O}\pa{m^{-2}}:&\!\!\!\!\!\!\!\!\!\!\!\!\quad \frac{\hbar^2}{2} \mathcal{G}^{AB}\paq{-\frac{2}{\gamma^2}\frac{\partial \gamma}{\partial q^A}\frac{\partial \gamma}{\partial q^B}+\frac{1}{\gamma}\frac{\partial^2 \gamma}{\partial q^A \partial q^B}+\frac{2}{\hbar^2}\frac{\partial S_0}{\partial q^A}\frac{\partial \zeta}{\partial q^B}}+\av{\hat H_k}_2\nonumber\\
&&\!\!\!\!\!\!\!\!\!\!\!\!\quad=\frac{\hbar^2}{2}\mathcal{G}^{AB}\av{\frac{\partial^2}{\partial q^A\partial q^B}}_{k,0},\label{spOm-2}
\ea
where by $\av{H_k}_0$, $\av{\frac{\partial^2}{\partial q^A\partial q^B}}_{k,0}$ and $\av{H_k}_2$ we mean the corresponding contributions to $\mathcal{O}\pa{m^0}$ and $\mathcal{O}\pa{m^{-2}}$ respectively \footnote{Let us note that Eq. (\ref{spOm0}) does not contain any terms associated with non-adiabatic transitions (r.h.s. of Eq. (\ref{wdwGk})) due to quantum gravitational effects. The omission of such terms is essentially the lowest order BO approximation.}. From Eq. (\ref{mateq0}) one obtains:
\ba{wkbmatexp}
&\!\!\!\!\!\!\!\!\!\!\!\!\!\!\!\!\!\!\!\!\!\!\!\!\!\!\!\!\!\!\!\!\!\!\!\!\!\!\mathcal{O}\pa{m^0}:&\!\!\!\!\!\!\!\!\!\!\!\!\quad -\frac{i\hbar}{2}\frac{\partial^2 S_1}{\partial v_k^2}+\frac{1}{2}\pa{\frac{\partial S_1}{\partial v_k}}^2+\frac{1}{2}\omega_k^2v_k^2-\av{\hat H_k}_0-\nonumber\\
&&\!\!\!\!\!\!\!\!\!\!\!\!\quad i\hbar \mathcal{G}^{AB}\frac{\partial S_0}{\partial q^A}\pa{\frac{\partial \ln \gamma}{\partial q^B}+\frac{i}{\hbar}\frac{\partial S_1}{\partial q^B}}=0,\label{matOm0}\\
&\!\!\!\!\!\!\!\!\!\!\!\!\!\!\!\!\!\!\!\!\!\!\!\!\!\!\!\!\!\!\!\!\!\!\!\!\!\!\mathcal{O}\pa{m^{-2}}:&\!\!\!\!\!\!\!\!\!\!\!\!\quad-\frac{i\hbar}{2}\frac{\partial^2 \chi}{\partial v_k^2}+\frac{\partial S_1}{\partial v_k}\frac{\partial \chi}{\partial v_k}+\frac{\omega_k^2}{2}v_k^2-\av{H_k}_2+\hbar^2 \mathcal{G}^{AB}\frac{\partial \ln \gamma}{\partial q^A}\frac{\partial \ln \gamma}{\partial q^B}\nonumber\\
&&\!\!\!\!\!\!\!\!\!\!\!\!\quad +\mathcal{G}^{AB}\frac{\partial S_0}{\partial q^A}\frac{\partial \chi}{\partial q^B}=-\frac{\hbar^2}{2}\mathcal{G}^{AB}\left(\av{\frac{\partial^2}{\partial q^A \partial q^B}}_{k,0}-\frac{1}{\gamma}\frac{\partial^2\gamma}{\partial q^A \partial q^B}\right.\nonumber\\
&&\!\!\!\!\!\!\!\!\!\!\!\!\quad\left.-\frac{i}{\hbar}\frac{\partial^2S_1}{\partial q^A \partial q^B}+\frac{1}{\hbar^2}\frac{\partial S_1}{\partial q^A}\frac{\partial S_1}{\partial q^B}
\right).\label{matOm-2}
\ea
Let us first note that (\ref{spOm2}) is equal to (\ref{Om2}). Furthermore, on comparing (\ref{spOm0}) and (\ref{matOm0}) and eliminating $\av {\hat H_k}_0$, one obtains Eq. (\ref{Om0}). Analogously on using (\ref{spOm-2}) and (\ref{matOm-2}) and eliminating $\av {\hat H_k}_2$ and $\mathcal{G}^{AB}\av{\frac{\partial^2}{\partial q^A \partial q^B}}_{k,0}$ one obtains (\ref{Om-2}). The BO factorization (\ref{defPsik}) combined with the $m$ expansion (\ref{wkbsplit}) are thus equivalent to the expansion (\ref{Sexp}) performed in \cite{KieferdS}.\\
We observe that BKK in Ref. \cite{KieferdS} now demand:
\be{gammadef}
\!\!\!\!\!\!\!\!\!\!\!\!\mathcal{G}^{AB}\frac{\partial}{\partial q^A}\paq{\frac{1}{2\gamma^2}\frac{\partial S_0}{\partial q^B}}=\mathcal{G}^{AB}\paq{-\frac{1}{\gamma^2}\frac{\partial \ln\gamma}{\partial q^A}\frac{\partial S_0}{\partial q^B}+\frac{1}{2\gamma^2}\frac{\partial^2 S_0}{\partial q^A \partial q^B}}=0,
\ee
which, on comparing with Eq. (\ref{spOm0}), requires $\av{\hat H_k}_0=0$ and is in contrast with the Gaussian ansatz solution subsequently used by Ref. \cite{KieferdS} in their Section 5. Subsequently in Ref. \cite{KieferdS} conformal time is introduced by
\be{conftime}
\frac{\partial}{\partial \eta}\equiv \mathcal{G}^{AB}\frac{\partial S_0}{\partial q^A}\pd{q^B}.
\ee
On now defining $\psi_k^{(0)}=\gamma \re^{\frac{i}{\hbar}S_1}$ Eq. (\ref{Om-2}) becomes
\ba{matKarray}
&&\!\!\!\!\!\!\!\!\!\!\!\!\!\!\!\!\!\!\!\!\!\!\!\!\paq{\frac{\partial \zeta}{\partial \eta}-\hbar^2\mathcal{G}^{AB}\frac{\partial \ln \gamma}{\partial q^A}\frac{\partial \ln \gamma}{\partial q^B}+\frac{\hbar^2}{2\gamma}\frac{\partial^2\gamma}{\partial q^A \partial q^B}}+\frac{\partial \chi}{\partial \eta}+\hbar^2 \mathcal{G}^{AB}\frac{\partial \ln \psi_k^{(0)}}{\partial q^A}\frac{\partial \ln \gamma}{\partial q^B}\nonumber\\
&&\!\!\!\!\!\!\!\!\!\!\!\!\!\!\!\!\!\!\!\!\!\!\!\!-\frac{\hbar^2}{2\psi_k^{(0)}}\mathcal{G}^{AB}\frac{\partial^2 \psi_k^{(0)}}{\partial q_A\partial q_B}-i\hbar\frac{\partial \ln\psi_k^{(0)}}{\partial v_k}\frac{\partial \chi}{\partial v_k}-\frac{i\hbar}{2}\frac{\partial^2\chi}{\partial v_k^2}=0,\label{matK}
\ea
and on setting the term in the square bracket equal to zero \cite{Kiefer90} one obtains 
\ba{matK88}
\frac{\partial \chi}{\partial \eta}&=&\frac{1}{\psi_k^{(0)}}\left(-\hbar^2 \mathcal{G}^{AB}\frac{\partial \psi_k^{(0)}}{\partial q^A}\frac{\partial \ln \gamma}{\partial q^B}+\frac{\hbar^2}{2}\mathcal{G}^{AB}\frac{\partial^2 \psi_k^{(0)}}{\partial q_A\partial q_B}\right.\nonumber\\
&&\left.+i\hbar\frac{\partial \psi_k^{(0)}}{\partial v_k}\frac{\partial \chi}{\partial v_k}+\frac{i\hbar \psi_k^{(0)}}{2}\frac{\partial^2\chi}{\partial v_k^2}\right)
\ea
which is Eq. (88) of Ref. \cite{KieferdS}.
However, setting the square bracket in Eq. (\ref{matK}) equal to zero is in contrast with our Eq. (\ref{spOm-2}) (after the introduction of time). 

In contrast, using our Eq. (\ref{spOm-2}), Eq. (\ref{matK88}) is modified as follows:
\ba{matK2}
\!\!\!\!\!\!\!\!\!\!\!\!\!\!\!\!\!\!\!\!\!\!\!\!\frac{\partial \chi}{\partial \eta}&\!\!\!\!\!\!\!\!\!\!\!\!\!\!=&!\!\!\!\!\!\!\!\frac{1}{\psi_k^{(0)}}\left(-\hbar^2 \mathcal{G}^{AB}\frac{\partial \psi_k^{(0)}}{\partial q^A}\frac{\partial \ln \gamma}{\partial q^B}+\frac{\hbar^2}{2}\mathcal{G}^{AB}\frac{\partial^2 \psi_k^{(0)}}{\partial q_A\partial q_B}\right.\nonumber\\
&&\left.+i\hbar\frac{\partial \psi_k^{(0)}}{\partial v_k}\frac{\partial \chi}{\partial v_k}+\frac{i\hbar \psi_k^{(0)}}{2}\frac{\partial^2\chi}{\partial v_k^2}\right)+\av{\hat H_k}_2-\frac{\hbar^2}{2}\mathcal{G}^{AB}\av{\frac{\partial^2}{\partial q^A\partial q^B}}_{k,0}.
\ea
On introducing $\psi_k^{(1)}$ as in \cite{KieferdS} by
\be{time}
\frac{\partial \chi}{\partial \eta}=\frac{m^2\hbar}{2}\frac{\partial \ln \frac{\psi_k^{(1)}}{\psi_k^{(0)}}}{\partial \eta}=\frac{\psi_k^{(0)}}{\psi_k^{(1)}}\frac{m^2\hbar}{i}\pa{\frac{1}{\psi_k^{(0)}}\frac{\partial \psi_k^{(1)}}{\partial \eta}-\frac{\psi_k^{(1)}}{{\psi_k^{(0)}}^2}\frac{\partial \psi_k^{(0)}}{\partial \eta}}
\ee
and using Eq. (\ref{matOm0}) one finally obtains (setting $\av{\hat H_k}_0+m^{-2}\av{\hat H_k}_2=\av{\hat H_k}$ since we only keep terms to order $m^{-2}$):
\ba{matK3}
&&-i\hbar\frac{\partial \psi_k^{(1)}}{\partial \eta}-\av{\hat H_k}\psi_k^{(1)}+\hat H_k\psi_k^{(1)}+\frac{\hbar^2}{2m^2}\mathcal{G}^{AB}\av{\frac{\partial^2}{\partial q^A\partial q^B}}_{k,0} \psi_k^{(1)}\nonumber\\
&&+\frac{\hbar^2}{m^2}\frac{\psi_k^{(1)}}{\psi_k^{(0)}}\mathcal{G}^{AB}\frac{\partial \ln \gamma}{\partial q^A}\frac{\partial \psi_k^{(0)}}{\partial q^B}-\frac{\hbar^2}{2m^2}\frac{\psi_k^{(1)}}{\psi_k^{(0)}}\mathcal{G}^{AB}\frac{\partial^2\psi_k^{(0)}}{\partial q^A \partial q^B}=0,
\ea
instead of Eq. (90) of Ref. \cite{KieferdS}. The Eq. (90) in \cite{KieferdS} violates unitarity as BKK explicitly claim in their comments to it. In contrast (\ref{matK3}) is free of unitarity violating contributions. 
Indeed one may identify a Schr\"odinger wave function $\psi_k^S$ as in done in \cite{T}:
\be{Swavef}
\psi_k^S=\re^{-\frac{i}{\hbar}\int^\eta\av{\hat H_k}d\eta'}\psi_k^{(1)},
\ee
which can be seen to satisfy (remembering Eq. (\ref{matOm0}) and only keeping terms to $\mathcal{O}\pa{m^{-2}}$):
\ba{unicheck}
&&i\hbar \frac{\partial}{\partial \eta}\int {\psi_k^S}^*\psi_k^S dv_k=i\hbar \frac{\partial}{\partial \eta}\int {\psi_k^{(1)}}^*\psi_k^{(1)} dv_k\nonumber\\
&&=i\hbar\int\paq{ {\psi_k^{(1)}}^*\frac{\partial \psi_k^{(1)}}{\partial \eta}-{\rm c.c.}}dv_k=0.
\ea
Thus there is no violation of unitarity. This, as has been pointed out before for the homogeneous case in Ref. \cite{Bertoni}, is a consequence of the inclusion of back reaction terms.
\section{Loop Gravity Approach}
Loop cosmology has become a topic of considerable interest (see e.g. \cite{Barrau}). In \cite{Barrau} several approaches to quantum cosmology are examined, but not the quantum geometrodynamical one. Three leading approaches to LQG are illustrated and compared: the so-called the ``effective constraint'', the ``separate universe approximation'' and the ``hybrid quantisation'' approaches. It is only the last one \cite{lqc} that claims a relation to the BO approach and is the one we shall address in this paper. Here in contrast with our case, a three torus rather than a flat three space is considered. Further, unlike the traditional approach we have illustrated in the Introduction, loop gravity in not quantised canonically but is quantised through a so-called polymer representation. This essentially means that rather than postulating canonical commutation relations between a coordinate and its conjugate momentum, thus leading to a representation of the momentum as a derivative with respect to the coordinate, one replaces the conjugate momentum by translations (or a limit thereof) with respect to the coordinate. Let us observe that a naive application of polymer quantisation, for example, to a harmonic oscillator does not appear to lead to the usual spectrum but rather leads to a band structure \cite{band}\footnote{Clearly analogous considerations would also apply to the polymer quantisation of, say, a massive scalar field which consists of an infinite set of oscillators.}. Moreover starting from LQG the semiclassical limit to the usual general relativity is rather problematic \cite{Nicolai}. 

In such a loop space formulation for gravitation one also introduces matter through a massive scalar field. Again one allows for non-homogeneous perturbations for both matter and geometry retaining terms at most to the second order (quadratic). The linear perturbative constraints relate the matter and geometric perturbations leading to the introduction of Mukhanov-Sasaki (MS) variables and a corresponding Hamiltonian formulation. The homogeneous matter is canonically quantised and a Fock space representation is introduced for the inhomogeneous modes. Such an approach is termed, by the authors, ``hybrid loop quantum cosmology''. In this approach also what the authors call a BO approach is followed and an ansatz is made to describe states of the system by a wave function decomposed as 
\be{lqcdec}
\Psi=\Gamma(\alpha,\phi)\psi(v,\phi)
\ee 
where $\alpha\equiv \ln \pa{a/a_0}$ is the logarithm of the FRW scale factor, $\phi$ is the homogeneous part of the matter field and $v$ is the MS variable. It is then assumed that an initial state $\Gamma\pa{\alpha,\phi_0}$ is essentially evolved by the square root of the lowest order homogenous matter Hamiltonian. One further assumes that the state $\Gamma\pa{\alpha,\phi_0}$ of the FRW geometry is so peaked on some value of $\phi$ that the corresponding state $\Gamma(\alpha,\phi)$ remains peaked for all considered values of $\phi$. In this approach the zero mode of the scalar field is interpreted as an internal time and no classical limit is taken for the gravitational part. 

After having introduced the ``BO ansatz'' through (\ref{lqcdec}) and the approximation that the state $\Gamma$ remains highly peaked on the operator values that encode the effect of the FRW geometry in the zero mode Hamiltonian constraint, effective classical equations for the MS variables are obtained. In the effective dynamics the perturbations are treated as classical, namely the creation and annihilation operators for the inhomogeneities are replaced by their classical counterparts. \\
The scope of the above discussion was just to illustrate the ``hybrid quantisation'' scheme in order to point out that it is does not involve a BO approach and how our first, true BO approach, is totally diverse with respect to it and is based on different physical principles. In particular in the former the gravitational part is polymer quantised whereas the homogeneous part of the scalar field $\phi$ is canonically quantised and they are both included in the same wavefunction (in contrast with the spirit of the BO approximation) which is assumed to be peaked on some value of $\phi$ which is related to the time. In our approach, even on limiting ourselves to the homogeneous case one can perform a BO factorisation separating gravitation and matter fields, introduce time through the semiclassical limit for the gravitational wave function and study inflation even for a non-classical homogeneous scalar field 
states \cite{Finelli}.

Our second approach, which we briefly illustrated in the Introduction, however is not a BO based one and bears some resemblance to the above loop space formulation, insofar as it does begin from a highly peaked homogeneous scalar field state and also involves a non-classical gravitational state. Let us examine it in more detail \cite{TVV}.\\
We start from the quantum (WDW) Einstein equation for a flat FRW universe in the coordinate ($a$, $\phi$) representation:
\be{wdwT}
\!\!\!\!\!\!\!\!\!\!
\!\!\!\!\!\!\!\!\!\!
\!\!\!\!\!\!\!\!\!\!
\!\!\!\!\!\!\!\!\!\!
\!\!\!
\pa{\frac{\hbar^2}{2m^2}\frac{\partial^2}{\partial a^2}\frac{1}{a}-\frac{\hbar^2}{2a^3}\frac{\partial^2}{\partial \phi^2}+\frac{\mu^2a^3}{2}\phi^2}\Psi_0(a,\phi)=\pa{\frac{\hbar^2}{2m^2}\frac{\partial^2}{\partial a^2}\frac{1}{a}+\hat H_I}\Psi_0(a,\phi)=0
\ee 
where, as before, $m$ is the Planck mass, $\mu$ is the scalar field (inflaton) mass, for convenience we have taken a particular ordering for the first term on the l.h.s. of (\ref{wdwT}) and $\Psi_0(a,\phi)$ is the total scale factor-inflaton homogeneous wave function. Here we shall be interested in configurations with large $a$, typical of a situation in an advanced stage of the inflationary phase, thus we are far away from any bounce or initial singularity, and one can verify that the difference between diverse factor orderings are irrelevant (negligible) in such a limit. A suitable choice of the matter part of the total wave function will lead to inflation: for example for $\phi$ constant one has a de Sitter universe. It is clear that for a positive definite inflaton Hamiltonian the wave function $\Psi_0$ will tend to have a strongly oscillatory dependence on $a$. The oscillation period will be so small that natural scales for the matter dynamics in $a$ are much longer and a coarse graining which can be, for example, chosen as an averaging over a period of oscillation in $a$, will be necessary to study the effective matter dynamics. A statistical interpretation as given in studies of the 
probability distributions in the hydrogen atom, as mentioned in the introduction, gives further insight \cite{rowe}. 

It is difficult to find a solution to Eq. (\ref{wdwT}) however following our previous work \cite{TVV} we solve it by factorising 
\be{expun}
\Psi_0(a,\phi)=a \psi(a)u(a,\phi)
\ee
and on making an ansatz for $u(a,\phi)$, show that it leads to an approximate solvable differential equation of $\psi(a)$. Let us observe that here $\Psi_0$ is chosen so that it is $0$ for $a\le 0$ as expected although the choice is unimportant since we shall obtain a solution in the inflationary, $a$ large, limit.\\
On observing that the homogeneous inflaton Hamiltonian is that of an $a$ dependent harmonic oscillator which can be cast in the usual form $\hat H_I=\hbar \mu \pa{\hat b^\dagger \hat b+\frac{1}{2}}$ with 
\be{bdef}
b=\sqrt{\frac{\mu a^3}{2\hbar}}\left(\hat\phi+\frac{i\hbar}{\mu a^3}
\hat\pi_{\phi}\right)
\ee
we make the following ansatz for $u(a,\phi)$
\be{gaussianu}
u(a,\phi)=\left(\frac{\mu a^3}{\pi\hbar}\right)^{\frac{1}{4}}
\exp\left[-\frac{\mu a^3}{2\hbar}(\phi-\phi_a)^2\right]
\ee
corresponding to the coherent state defined by $\hat b |\alpha\rangle=\alpha(a)|\alpha\rangle$ with $u\equiv \langle \phi|\alpha\rangle$, $\alpha(a)=\sqrt{\frac{\mu\phi_a^2a^3}{2\hbar }}$ and $\phi_a$ is a function of $a$. The expression (\ref{gaussianu}) is a Gaussian peaked around the $\phi_a$ with a width which decreases as the volume $a^3$ increases (we allow $\phi_a$ to have a small dependence on $a$ and to decrease for $a$ increasing). In particular for the case of chaotic inflation, in the slow roll approximation, the dependence is logarithmic \cite{marozzi}. 
Let us observe that if $\phi_a$ is a constant (which is the case we have previously considered) in the inflationary $a$ large limit we have a de Sitter universe. On allowing a small $a$ dependence for $\phi$ we are allowing a small deviation from de Sitter inflation corresponding to small non-zero slow roll parameters or if we wish a cosmological constant slowly varying with $a$. We shall return to this point at the end of this section.
If we now evaluate the following four diverse contributions to the WDW equation (\ref{wdwT}) obtained from the factorised wave function (\ref{expun}) 
\be{PDE}
u\:\partial_{a}^2 \psi+ 2\: \partial_a \psi_{a}u
\:\partial_{a}\psi+\psi\:\partial_{a}^2 u+2 \frac{m^2}{\hbar^{2}} a\,\psi\,\hat H_I u=0,
\ee
where $\partial_a\equiv\frac{\partial}{\partial a}$, $u(a,\phi)$ is given by (\ref{gaussianu}), and one finds that such contributions are strongly peaked on a small interval around $\phi_a$. Such an interval decreases as $a$ increases. Furthermore one has
\be{du}
\partial_{a} u(a,\phi)=\left[\frac{3}{4 a}-\frac{3 \mu a ^2}{2\hbar}
(\phi-\phi_{a})^2+\frac{\mu a^2}{\hbar}\pa{\phi-\phi_a}\phi_a\de{1}\right]u(a,\phi)\,,
\ee
with $\de{1}\equiv a\,\pa{\partial_a\phi_a} \phi_a^{-1}$,
\ba{ddu}
\!\!\!\!\!\!\!\!\!\!
\!\!\!\!\!\!\!\!\!\!
\!\!\!\!\!\!\!\!\!\!
\partial_{a}^2 \:u(a,\phi)=&&\left[-\frac{3}{16 a^2}-\frac{21 \mu a}{4\hbar}
(\phi-\phi_{a})^2+\frac{9 \mu^2 a^4}{4\hbar^2}(\phi-\phi_{a})^4\right.\nonumber\\
&&
\!\!\!\!\!\!
\!\!\!\!\!\!\!\!\!\!\left.\phantom{\frac{A}{B}}+\frac{2\mu a}{\hbar}\pa{\phi-\phi_a}\pa{13+2\de{2}}\de{1}\phi_a+\frac{\mu^2a^4}{\hbar^2}\pa{\phi-\phi_a}^2\de{1}\phi_a^2\right.\nonumber\\
&&
\!\!\!\!\!\!
\!\!\!\!\!\!\!\!\!\!\left.\phantom{\frac{A}{B}}-\frac{\mu^2a^4}{3\hbar^2}\pa{\phi-\phi_a}^3\de{1}\phi_a
\right]u(a,\phi)\,,
\ea
with $\de{2}\equiv a\pa{\partial_a \de{1}}\de{1}^{-1}$, and 
\be{dphiu}
\frac{\partial^2}{\partial\phi^2}u(a,\phi)=
\left[-\frac{\mu a^3}{\hbar}+\frac{\mu^2 a^6}{\hbar^2}
(\phi-\phi_{a})^2\right]u(a,\phi)\,.
\ee
Contributions of the form $x^n{\rm e}^{-x}$ which are recurrent in the above expressions are peaked in $x=n$ and in particular
\be{damp}
\!\!\!\!\!\!\!\!\!\!
\!\!\!\!\!\!\!\!\!\!
\!\!\!\!\!\!\!\!\!\!
{\rm max}_\phi\pag{\paq{\frac{\mu a^3}{\hbar}\pa{\phi-\phi_a}^2}^n u}=\pa{\frac{\mu a^3}{\pi \hbar}}^{1/4}\pa{\frac{2n}{{\rm e}}}^n=\pa{\frac{2n}{{\rm e}}}^n{\rm max}_\phi u
\ee
with the maximum at $\phi^{(n)}=\phi_a+\sqrt{\frac{2n\hbar }{\mu a^3}}$. Such maxima are slightly displaced w.r.t. to $\phi_a$ (where the maximum of $u(a,\phi)$ is located) but for large $a^3$ they are very close to it. For $a$ large the contributions to (\ref{du}), (\ref{ddu}) and (\ref{dphiu}) which are independent of $\de{1}$ and ${\de{2}}$ are of the same order of magnitude as $u\sim \paq{\mu a^3\pa{\phi-\phi_a}^2/\hbar}^n u$. On the other hand the remaining contributions to (\ref{du}), (\ref{ddu}) and (\ref{dphiu}) which contain $\de{1}$ and ${\de{2}}$ depend on the choice of $\phi_a$.

Let us note that, for $a$ large, the contribution of the inflaton potential to (\ref{PDE}) is strongly peaked around $\phi_a$ and can be approximated by
\be{infpot}
\frac{m^2}{\hbar^2}\mu^2a^4\phi^2 u(a,\phi)\simeq \frac{m^2}{\hbar^2}\mu^2a^4\phi_a^2 u(a,\phi).
\ee
On then assuming that $\phi_a$ is slowly varying w.r.t. $a$ (which is a reasonable assumption during inflation, at least at the classical level), the contributions (\ref{du}), (\ref{ddu}) and (\ref{dphiu}) to (\ref{PDE}) are negligible compared to (\ref{infpot}) and Eq. (\ref{PDE}) thus simplifies to
\be{graveq}
\partial_a^2\psi(a)+\frac{m^2}{\hbar^2}\mu^2\phi_a^2a^4\psi(a)=0.
\ee 
This last equation governs the quantum evolution of the scale factor during inflation and leads to strongly oscillatory dependence on $a$. We immediately see that for $\phi_a$ constant the second term in it corresponds to a cosmological constant. Such a solution is completely different to that obtained through a semiclassical approach to the gravitational wave equation obtained after the BO decomposition. Let us note that a different operator ordering in the kinetic term for gravity in (\ref{wdwT}) would lead to the same gravitational equation (\ref{graveq}). 

We have previously discussed the solutions obtained for Eq. (\ref{graveq}) for $\phi_a$ constant. We observe that different initial conditions can be taken for the solutions as suggested by Vilenkin \cite{Vilenkin} or Hartle and Hawking \cite{Hartle}. Coarse graining, corresponding to averaging over an oscillatory period for the gravitational wave function, leads to the introduction of time and evolution \cite{TVV}. On then allowing $\phi_a$ to have a small dependence on $a$ and adding cosmological perturbations one may determine the power spectrum and obtain predictions for large scale structures from such an approach \cite{WIP}.

\section{Conclusions}
The purpose of this paper was to illustrate and compare diverse approaches to the cosmology of the matter-gravity system all of which have the ultimate aim of finding expressions and corrections for the power spectrum generated by inflation and refer to the BO method. As a basis for comparison we use the approach we have previously studied in detail \cite{T} and which we may term a traditional BO approach and hinges on the fact that the Planck mass is much greater than the usual matter mass. This allows us to separate the global matter gravity wave function into a purely gravitational part, wherein gravitation is driven by the mean matter Hamiltonian, together with an additional back-reaction due to transitions between diverse matter states induced by the variation of the FRW radius, and a non linear matter equation which again includes a back- reaction due to transition between matter states. All of this in the semiclassical limit for gravitation.

Let us better illustrate what we mean by the above mentioned transitions. To lowest order in the BO approach generally such transitions are neglected. This means gravitation is just driven by the mean matter Hamiltonian. Matter evolution, on the other hand, to lowest order and for each mode, is essentially described by a harmonic oscillator-type Hamiltonian with a time dependent frequency (MS equation). One can solve this equation, formally exactly, for the corresponding states \cite{Lewis}. The corrections to both the gravitational and the matter evolution equations are associated with transitions between such eigenstates \cite{Bertoni}.

The alternative approaches we addressed for the comparison are two which claim to use, or be inspired by, a BO approach but actually differ significantly from the beginning insofar as the homogeneous part of the matter (scalar) field is lumped into the ``heavy'' part of the factorised wave function together with the homogeneous gravitational part and the two are not separated in contrast with our traditional approach (unless, of course, the scalar field has a Planck mass)

The first (BKK) approach for our comparison is quantum geometrodynamical \cite{KieferdS}, being based on the WDW equation, is clearly related to our method and also involves a semiclassical approximation. We have followed for it a treatment analogous to the one employed in our traditional BO approach. This has allowed us to separate the diverse equations obtained on expanding in powers of $m$ (Planck mass) for both matter and gravitation. One finds corrections to the results previously obtained due to the presence of back-reaction terms involving the average of the matter Hamiltonian and the homogeneous gravitational kinetic terms with respect to the matter wave function. As a consequence of the presence of such terms no violation of unitarity is found. A similar result has been obtained previously for the purely homogeneous case \cite{Bertoni}.

The only loop quantum cosmology approach that refers to BO is the so-called ``hybrid quantisation'' one and differs completely from our traditional BO method. A comparison between the two approaches is difficult since the former is also based on a different quantisations for matter (canonical quantisation) and gravitation (polymer quantisation for the homogeneous part). Furthermore time is introduced, not through a semiclassical limit for gravitation (which remains quantised in contrast with our approach), but through the interpretation of the homogeneous part of the scalar field as a relational time. This is very different both technically and physically from our traditional BO approach.

On the other hand another approach we have previously considered \cite{TVV} bears some resemblance to the above loop space approach. In such an approach we studied the quantum minisuperspace homogeneous-inflaton system in the inflationary, a large, limit. Beginning with a suitable ansatz for the initial matter (inflaton) state (a highly peaked normal distribution) we solved the WDW equation for the gravitational wave function finding it to have a highly oscillatory behaviour. On introducing normal matter (or cosmological fluctuations) and coarse graining for the gravitational wave function (averaging over a fluctuation period) one obtains time and evolution. Thus, analogously with the loop space case, we have a quantum gravitational state and a highly peaked scalar (inflaton) field. Here no classical limit is taken for the gravitational wave function and it will be interesting to study cosmological perturbations in such a context and the predictions for the CMB signatures.

The results of a quantum geometrodynamical and loop space formulations have been examined before \cite{Calcagni}. However the cases examined are the loop space ``effective constraint'' approach, which does not refer to BO and can lead to a power enhancement on large scales, and the geometrodynamical approach \cite{KieferdS} which we disagree with. In any case our emphasis has been on approaches which claim a BO link and not to review different approaches or fits to the data (which few have done).

The work of A.K. was partially supported by the RFBR grant No 17-02-01008.


\end{document}